\documentclass[aps,prd,twocolumn,superscriptaddress,showpacs]{revtex4}
\usepackage{graphicx}
\usepackage{lscape}
\usepackage{epsfig,epsf}
\usepackage{amsmath}
\usepackage{amsthm}
\usepackage{amsfonts}
\usepackage{amssymb}
\usepackage{dsfont}
\usepackage{multirow}
\usepackage{appendix}
\usepackage{slashed}
\usepackage[active]{srcltx}
\usepackage{psfrag}
\usepackage{multirow}

\newcommand{\be}{\begin{equation}}
\newcommand{\ee}{\end{equation}}
\newcommand{\bea}{\begin{eqnarray}}
\newcommand{\eea}{\end{eqnarray}}
\newcommand{\nn}{\nonumber}

\def\R1{\varepsilon_1}
\def\E8{\varepsilon_8}

\def\ga{\gamma}

\def\s1{\hat s}
\def\ds{\displaystyle}

\newcommand{\bd}{\begin{displaymath}}
\newcommand{\ed}{\end{displaymath}}

\newcommand{\f}{\frac}

\def\R1{\varepsilon_1}
\def\E8{\varepsilon_8}

\def\ga{\gamma}

\def\ds{\displaystyle}
\def\beq{\begin{equation}}
\def\eeq{\end{equation}}
\def\bea{\begin{eqnarray}}
\def\eea{\end{eqnarray}}
\def\beeq{\begin{eqnarray}}
\def\eeeq{\end{eqnarray}}

\def\vel{\left|}
\def\ver{\right|}
\def\nnb{\nonumber}
\def\ga{\left(}
\def\dr{\right)}

\def\rar{\rightarrow}
\def\nnb{\nonumber}

\def\ba{\begin{array}}
\def\ea{\end{array}}

\def\xis0{{\Xi^{*0}}}

\def\g5{\gamma_5}

\def\es{\!\!\! &=& \!\!\!}

\def\ar{&+& \!\!\!}
\def\ek{&-& \!\!\!}

\begin{document}

\title{Impact of scalar leptoquarks on heavy baryonic decays}
\date{\today}
\author{K.~Azizi}
\affiliation{Department of Physics, Do\v{g}u\c{s} University, Ac{\i}badem-Kad{\i}k\"{o}y, 34722
\.{I}stanbul, Turkey}
\author{A.~T.~Olgun}
\affiliation{Vocational School Kad{\i}k\"oy Campus, Okan University,
Hasanpa\c sa-Kad{\i}k\"oy, 34722 \.{I}stanbul, Turkey}
\author{Z.~Tavuko\u glu}
\affiliation{Vocational School Kad{\i}k\"oy Campus, Okan University,
Hasanpa\c sa-Kad{\i}k\"oy, 34722 \.{I}stanbul, Turkey}

\begin{abstract}
We present a study on the impact of scalar leptoquarks on the semileptonic decays of $ \Lambda_b$, 
$\Sigma_b $ and $\Xi_b $. To this end, we calculate the differential 
branching ratio and lepton forward-backward asymmetry defining  the processes $ \Lambda_b \rightarrow \Lambda \ell^+  \ell^-$, $\Sigma_b  \rightarrow \Sigma \ell^+  \ell^-$ 
and $\Xi_b \rightarrow \Xi \ell^+  \ell^-$, with $\ell$ being $\mu$ or $\tau$,
 using the form factors calculated via light cone QCD
in full theory. In  calculations, the errors of form factors are  taken into account. We compare the results obtained in leptoquark model with those of the standard model as well as the existing lattice QCD predictions and experimental data. 
\end{abstract}


\maketitle

\section{Introduction}
\label{Sec:Int}
The physics of transitions based on $ b \rightarrow s \ell^+  \ell^-$ at quark level constitutes one of the main directions of the research in high energy 
and particle physics both theoretically and experimentally as new physics effects can contribute to such decay channels.
The flavor changing neutral current (FCNC) transitions of $ \Lambda_b \rightarrow \Lambda \ell^+  \ell^-$, $\Sigma_b  \rightarrow \Sigma \ell^+  \ell^-$ 
and $\Xi_b \rightarrow \Xi \ell^+  \ell^-$  are among important baryonic decay channels that can be used as sensitive probes to indirectly search for new physics contributions.
Especially, the rare $ \Lambda_b \rightarrow \Lambda \ell^+  \ell^-$ decay  channel has been in the focus of much attention in recent years both theoretically and experimentally.
 The first measurement on the $ \Lambda_b \rightarrow \Lambda \mu^+  \mu^-$ process
 has been reported by the CDF Collaboration \cite{Aaltonen:2011qs} with $24$ signal events and a statistical significance of $5.8$ Gaussian standard deviations.
 Using the $p\overline{p}$ collisions data samples corresponding
to $6.8 fb^{-1}$ and $\sqrt{s}=1.96$ TeV collected by the CDF II detector, the differential branching ratio for  the $ \Lambda_b \rightarrow \Lambda \mu^+  \mu^-$ decay channel has been measured to be   $dBr(\Lambda _b^0 \rightarrow \Lambda \mu^+ \mu^-)/dq^2=[1.73\pm0.42(stat)\pm0.55(syst)]\times10^{-6}$ \cite{Aaltonen:2011qs}.
 The differential branching fraction of $\Lambda _b^0 \rightarrow \Lambda \mu^+ \mu^-$ decay channel has also been  measured as $dBr (\Lambda _b^0 \rightarrow \Lambda \mu^+ \mu^-)/dq^2 = (1.18 \;^{+\,0.09}_{-\,0.08} \pm 0.03 \pm 0.27 )
\times 10^{-7}$ GeV$^2/$c$^4$ at $15 $ GeV$^2/$c$^4\leq$   $q^2 \leq 20$  GeV$^2/$c$^4$ region  by the LHCb Collaboration \cite{LHCb}. The LHCb Collaboration has also measured  the lepton  forward-backward asymmetries associated to this transition  as $ A_{FB}^{\mu}=-0.05 \pm 0.09 (stat)\pm 0.03 (syst)$ at $15 $ GeV$^2/$c$^4\leq$   $q^2 \leq 20$  GeV$^2/$c$^4$ region \cite{LHCb}. The order of branching ratio in $ \Lambda_b \rightarrow \Lambda e^+  e^-$, $ \Lambda_b \rightarrow \Lambda \tau^+  \tau^-$ as well as in $\Sigma_b  \rightarrow \Sigma \ell^+  \ell^-$ and $\Xi_b \rightarrow \Xi \ell^+  \ell^-$ (for all leptons) indicates that these channels are all
accessible at LHC (for details see Refs.\cite{Aliev:2010uy,Azizi:2011if,Azizi:2011mw,Azizi:2012yg,Azizi:2013eta}) . We hope with the RUN II data at the center of mass energy 13 TeV  it will be possible to measure different physical quantities related to these FCNC loop level rare transitions in near future.

The LHC RUN II may provide opportunities to search for various new physics scenarios. One of the important new physics models that has been proposed to overcome the problems of  some inconsistencies
 between the SM predictions and experimental data, is the leptoquark (LQ) model. Hereafter, by LQ model we mean a minimal renormalizable scalar leptoquark model which will be explained in some details in next section.  As an example for the LHC constraints and prospects for scalar leptoquarks explaining 
the $\overline{B}  \rightarrow D^{(*)} \tau \overline{\nu}$ anomaly see \cite{Dumont:2016xpj}. LQs are hypothetical  color triplet bosons that couple to 
 leptons and  quarks \cite{pdg}. LQs carry both baryon (B) and lepton (L) quantum numbers with color and electric charges. The spin number of a leptoquark state can be  $0$ or $1$,
 corresponding to a scalar leptoquark or vector leptoquark. If the leptoquarks violate both the baryon and lepton numbers, they are generally considered to be heavy particles at the level of 
${\cal O}( 10^{15})$ GeV in order to prevent the proton decay. For more detailed information about leptoquark models and the recent experimental and theoretical progresses, see \cite{Dorsner:2016wpm,Davidson:1993qk,Hewett:1997ce,Shanker:1981mj,Fajfer:2008tm,Saha:2010vw,Davidson:2011zn,Kosnik:2012dj,Arnold:2013cva,Sakaki:2013bfa,Mohanta:2013lsa,Allanach:2015ria,Sahoo:2015wya,Sahoo:2015qha,Sahoo:2015fla,Sahoo:2015pzk,Kumar:2016omp,Dorsner:2009cu,Dorsner:2011ai,Becirevic:2016oho,Becirevic:2016yqi}. 

In the light of progresses about LQs, we calculate the differential branching ratio and lepton forward-backward asymmetry corresponding to the  $ \Lambda_b \rightarrow \Lambda \ell^+  \ell^-$, $\Sigma_b  \rightarrow \Sigma \ell^+  \ell^-$ and $\Xi_b \rightarrow \Xi \ell^+  \ell^-$ processes in a scalar LQ model.  In the calculations, we use the form factors as the main inputs calculated from the light cone QCD sum rules in full theory. We also encounter the errors of the form factors to the calculations. We compare the regions swept by the LQ model with those of the SM and search for deviations of the LQ model predictions with those of the SM. We also compare the results with the available lattice predictions and experimental data.

The outline of this article is as follow. In next section, we present the effective Hamiltonian 
responsible for the transitions under consideration both in the SM and LQ models. In section III, we present the transition amplitude and matrix elements defining the above transitions. In section IV, we calculate the differential decay rate and the lepton forward-backward asymmetry  in the baryonic  $ \Lambda_b \rightarrow \Lambda \ell^+  \ell^-$, $\Sigma_b  \rightarrow \Sigma \ell^+  \ell^-$ and $\Xi_b \rightarrow \Xi \ell^+  \ell^-$ channels and numerically analyze  the results obtained. We compare the LQ predictions with those of the SM and existing  lattice results and experimental data also in this section.
\section{The Effective Hamiltonian and Wilson Coefficients}
At the quark level the  effective Hamiltonian, defining the above mentioned $b \rar s \ell^+ \ell^- $ based transitions,  in terms of Wilson coefficients and different operators 
  in   SM is generally  defined  as  \cite{Buchalla:1995vs,Altmannshofer:2008dz}
\begin{eqnarray} \label{HeffSM} 
{\cal H}^{eff}_{SM} &=& {G_F \alpha_{em} V_{tb}
V_{ts}^\ast \over 2\sqrt{2} \pi} \Bigg[ C^{eff}_{9}
\bar{s}\gamma_\mu (1-\gamma_5) b \, \bar{\ell} \gamma^\mu \ell \nnb \\
&+&C^{\prime~eff}_{9}
\bar{s}\gamma_\mu (1+\gamma_5) b \, \bar{\ell} \gamma^\mu \ell \nnb \\
 &+& C_{10}  \bar{s} \gamma_\mu (1-\gamma_5) b \, \bar{\ell}
\gamma^\mu
\gamma_5 \ell \nnb \\
 &+& C^{\prime}_{10}  \bar{s} \gamma_\mu (1+\gamma_5) b \, \bar{\ell}
\gamma^\mu
\gamma_5 \ell \nnb \\
&-&  2 m_b C^{eff}_{7} {1\over q^2} \bar{s} i \sigma_{\mu\nu} q^{\nu}
(1+\gamma_5) b \, \bar{\ell} \gamma^\mu \ell 
\nnb \\
&-&  2 m_b C^{\prime~eff}_{7} {1\over q^2} \bar{s} i \sigma_{\mu\nu} q^{\nu}
(1-\gamma_5) b \, \bar{\ell} \gamma^\mu \ell 
\Bigg]~, 
\end{eqnarray}
where $G_{F}$ is the Fermi weak coupling constant, $\alpha_{em}$ is the fine structure constant at $Z$ mass scale, $V_{tb}$ and
$V_{ts}^\ast$ are elements of the Cabibbo-Kobayashi-Maskawa (CKM) matrix, the $C^{(\prime)~eff}_{9}$, $C^{(\prime)}_{10}$ and $C^{(\prime)~eff}_{7}$ are the SM Wilson coefficients and $q^2$ is the transferred momentum squared. Here the superscript ``eff" refers to the shifts in the corresponding coefficients due to the effects of four-quark operators at large $ q^2 $.  The primed coefficients are ignored since the Hamiltonian does not receive any contribution from the  corresponding operators in the SM. We collect the explicit expressions of the   Wilson coefficients $C^{eff}_{9}$, $C_{10}$ and $C^{eff}_{7}$  in the Appendix: A.

Considering the additional contributions arising from the exchange of scalar leptoquarks, 
the effective Hamiltonian is modified. The modified Hamiltonian in LQ model is obtained from Eq. (\ref{HeffSM}) by the replacements 
$ C^{eff}_{9}\rightarrow  C^{eff,tot}_{9}$, $ C^{\prime~eff}_{9}\rightarrow  C^{\prime~eff,tot}_{9}$, $ C_{10}\rightarrow  C^{tot}_{10}$ and $ C^{\prime}_{10}\rightarrow  C^{\prime ~tot}_{10}$.
Here,  $C^{eff,tot}_{9}$, $C^{\prime~eff,tot}_{9}$, $C^{tot}_{10}$ and $C^{\prime~tot}_{10}$, with the superscript ``tot" being referring to ``total" , are new Wilson coefficients. These coefficients contain contributions from both the SM and LQ models. Note that the Wilson coefficients  $C^{eff}_{7}$ and $C^{\prime~eff}_{7}$ remain unchanged compared to the SM.  The new Wilson coefficients are given as (for details  see  for instance \cite{Mohanta:2013lsa,Sahoo:2015wya,Sahoo:2015qha,Sahoo:2015pzk,Sahoo:2015fla})
\begin{eqnarray}
C^{eff,tot}_{9} &=& C^{eff}_{9}  + C^{LQ}_{9}~,  \nnb \\
C^{\prime~eff,tot}_{9} &=&C^{\prime~eff}_{9}+ C^{\prime~LQ}_{9}~,  \nnb \\
C^{tot}_{10} &=& C_{10}  + C^{LQ}_{10}~,  \nnb \\
C^{\prime~tot}_{10} &=& C^{\prime}_{10}+C^{\prime~LQ}_{10},
\end{eqnarray}
where the coefficients $C^{LQ}_{9}$ and $C^{LQ}_{10}$ receive contributions from the exchange of the scalar leptoquarks $X^{(7/6)}=(3, 2, 7/6)$ but the primed Wilson coefficients $C^{\prime~LQ}_{9}$ and $C^{\prime~LQ}_{10}$  pick up contributions from the exchange  of the  scalar leptoquarks  $X^{(1/6)}=(3, 2, 1/6)$.  Here we should remark that we consider the effects of the  above two scalar  leptoquarks on the Wilson coefficients since this representation does not allow proton decay at tree-level. We do not consider the effects of the vector leptoquarks on the processes under consideration. Hence, in the present study we consider the minimal renormalizable scalar leptoquark models including one single additional representation of $ SU(3) \times SU(2)\times U(1) $ which guarantees that the proton does not decay. This requisite can only be satisfied by the models that have the representation of $X^{(7/6)}=(3, 2, 7/6)$ and $X^{(1/6)}=(3, 2, 1/6)$ scalar leptoquarks under the above gauge group (for details see  for instance \cite{Arnold:2013cva,Sahoo:2015wya}).

 Thus the    coefficients $C^{LQ}_{9}$ and $C^{LQ}_{10}$   are obtained as \cite{Mohanta:2013lsa,Sahoo:2015wya,Sahoo:2015qha,Sahoo:2015pzk,Sahoo:2015fla}
\begin{eqnarray}
C^{LQ}_{9}= C^{LQ}_{10} = - \frac{ \pi}{2 \sqrt 2 G_F \alpha_{em}  V_{tb} V_{ts}^* }\frac{\lambda_e^{23} \lambda_e^{22 *}}{
M_Y^2}\;,\label{C10LQ}
\end{eqnarray}
and the primed Wilson coefficients    $C^{\prime~LQ}_{9}$ and $C^{\prime~LQ}_{10}$ are found as \cite{Mohanta:2013lsa,Sahoo:2015wya,Sahoo:2015qha,Sahoo:2015pzk,Sahoo:2015fla}
\begin{eqnarray}
C^{\prime~LQ}_{9}= - C^{\prime~LQ}_{10} = \frac{ \pi}{2 \sqrt 2 ~G_F \alpha_{em}  V_{tb}V_{ts}^*} \frac{\lambda_s^{22} \lambda_b^{32*}}{M_V^2}\;,\label{c10np1}
\end{eqnarray}
where $ Y $ and $ V $ are the two components of doublet LQ, $ X=(V_\alpha, Y_\alpha) $,  with  $ M_Y $ and ${M_V}$ being representing  the masses of the components of the scalar leptoquarks (for details on the LQ interaction Lagrangian and corresponding notations see \cite{Sahoo:2016nvx}). It is assumed that each individual leptoquark contribution to the branching ratio does not exceed the experimental result.  Here
\begin{eqnarray}\label{constraints}
0\leq\left|\frac{\lambda_{e}^{23} \lambda_e^{22*}}{M_Y^2}\right| = \left|\frac{\lambda_{s}^{22} \lambda_b^{32*}}{M_V^2}\right| \leq 5 \times 10^{-9} ~{\rm GeV}^{-2}\;,
\end{eqnarray}
 obtained via the fitting of the model parameters to the $ B_s \rightarrow \mu^+\mu^- $ data \cite{Sahoo:2016nvx}. In Eq. (5) we assummed that the contributions of the two components $ Y $ and $ V $ are equal.

\label{sec:Vertex}
\section{Transition amplitude and matrix elements}
 Generally, the amplitude of the transition responsible for the $ \Lambda_b \rightarrow \Lambda \ell^+  \ell^-$, $\Sigma_b  \rightarrow \Sigma \ell^+  \ell^-$ and $\Xi_b \rightarrow \Xi \ell^+  \ell^-$ 
baryonic decays is provided with sandwiching the  effective Hamiltonian between   the initial and final baryonic states,
\begin{eqnarray}\label{amplitude}
{\cal M}^{ {\cal B}_{Q}  \rightarrow {\cal B}  \ell^+ \ell^-} = \langle {\cal B} (p) \mid{\cal H}^{eff}\mid 
{\cal B}_{Q} (p+q,s) \rangle~,
\end{eqnarray}
where ${\cal B}$ represents  $\Lambda$, $\Sigma$ and $\Xi$ baryons and $ Q $ corresponds to $ b  $ quark.  
To get the transition amplitude,  we need to consider  the following transition matrix elements parametrized  in terms of twelve form factors in full QCD, i.e.,  without any expansion in the heavy quark mass or large hadron energies:
\begin{eqnarray}\label{SMtransmatrix} 
\langle {\cal B } (p) \mid \bar s \gamma_\mu (1&-&\gamma_5) b \mid {\cal B}_{Q} (p+q,s)\rangle = \nnb \\ 
&&\bar {u}_{\cal B} (p)\Bigg [\gamma_{\mu}f_{1}(q^{2}) +{i}\sigma_{\mu\nu}q^{\nu}f_{2}(q^{2}) \nnb \\
&&+ q^{\mu}f_{3}(q^{2}) - \gamma_{\mu}\gamma_5 g_{1}(q^{2}) \nnb \\
&&-{i}\sigma_{\mu\nu}\gamma_5q^{\nu}g_{2}(q^{2}) \nnb \\
&&- q^{\mu}\gamma_5 g_{3}(q^{2})  \vphantom{\int_0^{x_2}}\Bigg] u_{{\cal B}_{Q}}(p+q,s)~,\nnb \\
 \langle
{\cal B} (p) \mid  \bar s \gamma_\mu (1&+&\gamma_5) b \mid {\cal B}_{Q} (p+q,s)\rangle = \nnb \\
&&\bar {u}_{\cal B} (p) \Bigg[\gamma_{\mu}f_{1}(q^{2})+{i}\sigma_{\mu\nu}q^{\nu}f_{2}(q^{2}) \nnb \\
&&+ q^{\mu}f_{3}(q^{2}) + \gamma_{\mu}\gamma_5 g_{1}(q^{2}) \nnb \\
&&+{i}\sigma_{\mu\nu}\gamma_5q^{\nu}g_{2}(q^{2}) \nnb \\
&&+ q^{\mu}\gamma_5 g_{3}(q^{2}) \vphantom{\int_0^{x_2}}\Bigg] u_{{\cal B}_{Q}}(p+q,s)~,\nnb \\
\langle {\cal B} (p) \mid \bar s i \sigma_{\mu\nu}q^{\nu} (1&+& \gamma_5)b \mid {\cal B}_{Q} (p+q,s)\rangle =\nnb \\ 
&&\bar {u}_{\cal B} (p)\Bigg[\gamma_{\mu}f_{1}^{T}(q^{2})+{i}\sigma_{\mu\nu}q^{\nu}f_{2}^{T}(q^{2}) \nnb \\ 
&&+q^{\mu}f_{3}^{T}(q^{2}) + \gamma_{\mu}\gamma_5 g_{1}^{T}(q^{2}) \nnb \\
&&+{i}\sigma_{\mu\nu}\gamma_5q^{\nu}g_{2}^{T}(q^{2}) \nnb \\
&&+ q^{\mu}\gamma_5 g_{3}^{T}(q^{2})\vphantom{\int_0^{x_2}}\Bigg] u_{{\cal B}_{Q}}(p+q,s)~,\nnb \\
\langle {\cal B} (p) \mid \bar s i \sigma_{\mu\nu}q^{\nu} (1&-& \gamma_5) b \mid {\cal B}_{Q} (p+q,s)\rangle = \nnb \\ 
&&\bar {u}_{\cal B} (p) \Bigg[\gamma_{\mu}f_{1}^{T}(q^{2})+{i}\sigma_{\mu\nu}q^{\nu}f_{2}^{T}(q^{2}) \nnb \\ 
&&+q^{\mu}f_{3}^{T}(q^{2}) - \gamma_{\mu}\gamma_5 g_{1}^{T}(q^{2}) \nnb \\ 
&&-{i}\sigma_{\mu\nu}\gamma_5q^{\nu}g_{2}^{T}(q^{2}) \nnb \\
&&- q^{\mu}\gamma_5 g_{3}^{T}(q^{2}) \vphantom{\int_0^{x_2}}\Bigg] u_{{\cal B}_{Q}}(p+q,s)~,\nnb \\
\end{eqnarray}
where the $u_{{\cal B}_{Q}}$ and ${u}_{\cal B}$ represent spinors
of the initial and final states, respectively. The $f^{(T)}_i$ and $g^{(T)}_i$ ($i$ running from $1$ to $ 3$) are transition form factors .
The values of these form factors corresponding to  $ \Lambda_b \rightarrow \Lambda \ell^+  \ell^-$, $\Sigma_b  \rightarrow \Sigma \ell^+  \ell^-$ and $\Xi_b \rightarrow \Xi \ell^+  \ell^-$ 
transitions  and calculated via light cone sum rules in full theory   are taken from \cite{Aliev:2010uy}, \cite{Azizi:2011if} and \cite{Azizi:2011mw}, respectively (for form factors of $ \Lambda_b $ channel calculated with different phenomenological models see also for instance \cite{Feldmann:2011xf,Boer:2014kda,Wang:2015ndk}). These form factors are also available in lattice QCD in $ \Lambda $ channel \cite{Detmold:2016pkz}.

Using the  above   transition matrix elements in terms of form factors, we get the amplitude of the transitions 
 $ \Lambda_b \rightarrow \Lambda \ell^+  \ell^-$, $\Sigma_b  \rightarrow \Sigma \ell^+  \ell^-$ and $\Xi_b \rightarrow \Xi \ell^+  \ell^-$  in the SM and LQ  as
\begin{widetext}
\begin{eqnarray}\label{amplitude1}
{\cal M}_{SM}^{ {\cal B}_{Q} \rightarrow {\cal B} \ell^+ \ell^-} &=& {G_F \alpha_{em} V_{tb}V_{ts}^\ast \over 2\sqrt{2} \pi} 
\Bigg\{\Big[{\bar u}_{\cal B}(p) ( \gamma_{\mu}[{\cal A}_1^{SM} R + {\cal B}_1^{SM} L] + 
{i}\sigma_{\mu\nu} q^{\nu}[{\cal A}_2^{SM} R + {\cal B}_2^{SM} L] \nnb \\
&+& q^{\mu} [{\cal A}_3^{SM} R + {\cal B}_3^{SM} L]) 
u_{{\cal B}_{Q}}(p+q,s) \Big] \, (\bar{\ell} 
\gamma^\mu \ell)\nnb \\
&+& \Big[{\bar u}_{\cal B}(p) ( \gamma_{\mu}[{\cal D}_1^{SM} R + {\cal E}_1^{SM} L]+ {i}\sigma_{\mu\nu} q^{\nu}[{\cal D}_2^{SM} R 
+{\cal E}_2^{SM} L] \nnb \\
&+& q^{\mu} [{\cal D}_3^{SM} R + {\cal E}_3^{SM} L]) u_{{\cal B}_{Q}}(p+q,s) \Big] \,(\bar{\ell} 
\gamma^\mu \gamma_5 \ell) \Bigg\}, ~\nnb \\
\end{eqnarray}
and  
\begin{eqnarray}\label{amplitude2}
{\cal M}_{tot}^{ {\cal B}_{Q}  \rightarrow {\cal B}  \ell^+ \ell^-} &=& {G_F \alpha_{em} V_{tb}V_{ts}^\ast \over 2\sqrt{2} \pi} 
\Bigg\{\Big[{\bar u}_{\cal B}(p) ( \gamma_{\mu}[{\cal A}_1^{tot} R + {\cal B}_1^{tot} L] 
+ {i}\sigma_{\mu\nu} q^{\nu}[{\cal A}_2^{tot} R + {\cal B}_2^{tot} L] \nnb \\
&+& q^{\mu} [{\cal A}_3^{tot} R + {\cal B}_3^{tot} L]) u_{{\cal B}_{Q}}(p+q,s) \Big] \, (\bar{\ell} \gamma^\mu \ell)\nnb \\
&+& \Big[{\bar u}_{\cal B}(p)( \gamma_{\mu}[{\cal D}_1^{tot} R + {\cal E}_1^{tot} L]
+ {i}\sigma_{\mu\nu} q^{\nu}[{\cal D}_2^{tot} R +{\cal E}_2^{tot} L] \nnb \\ 
&+& q^{\mu} [{\cal D}_3^{tot} R + {\cal E}_3^{tot} L]) u_{{\cal B}_{Q}}(p+q,s) \Big] \,(\bar{\ell} \gamma^\mu \gamma_5 \ell) 
\Bigg\} ~,\nnb \\
\end{eqnarray}
\end{widetext}
where $R=(1+\gamma_5)/2$  and $L=(1-\gamma_5)/2$ and the calligraphic coefficients  are collected  in Appendix: B.
\section{Physical Observables}

\label{sec:Vertex}

In this section we would like to calculate some physical observables such as the differential decay width, the differential branching ratio and the lepton forward-backward asymmetry for the 
considered decay channels.

\subsection{The differential decay width}
Using the decay amplitudes and transition matrix elements in terms of form factors, we find the differential decay rate 
defining the transitions under consideration  in the LQ model as
\begin{eqnarray} \label{DDR}
\frac{d^2\Gamma_{tot}}{d\hat
sdz}(z,\hat s) &=& \frac{G_F^2\alpha^2_{em} m_{{\cal B}_Q}}{16384
\pi^5}| V_{tb}V_{ts}^*|^2 v \sqrt{\lambda(1,r,\hat s)} \, \nnb \\ 
&\Bigg[&{\cal
T}_{0}^{tot}(\hat s)+{\cal T}_{1}^{tot}(\hat s) z +{\cal T}_{2}^{tot}(\hat s)
z^2\Bigg]~, 
\nnb\\ 
\end{eqnarray}
where $v=\sqrt{1-\frac{4 m_\ell^2}{q^2}}$ is the lepton velocity, $\lambda=\lambda(1,r,\hat s)=(1-r-\hat s)^2-4r\hat s$ 
is the usual triangle function, $\hat s= q^2/m^2_{{\cal B}_Q}$, $r= m^2_{{\cal B}}/m^2_{{\cal B}_Q}$ and $z=\cos\theta$ with $\theta$ 
being the angle between momenta of the lepton $l^+$ and the  ${\cal B}_Q$ in the center of mass of leptons. The calligraphic
${\cal T}_{0}^{tot}(\hat s)$, ${\cal T}_{1}^{tot}(\hat s)$ and ${\cal T}_{2}^{tot}(\hat s)$ functions are given in Appendix: B. 

\subsection{The differential branching ratio}
Using the  expression of the differential decay width, in this subsection, we numerically analyze the differential branching ratio in terms of $q^2$ for the
decay channels under consideration. For this aim, we present the values of some input parameters and the quark masses in $\overline{MS}$ scheme 
used in the numerical analysis in tables 1 and 2 \cite{pdg}. 
Using the numerical values in these tables and  the expressions presented in the appendix A, we find the   values/intervals $C^{eff}_{7}=-0.295$,   $C^{eff}_{9}=[1.573,6.625]$,  $C_{10}=-4.260$,  $C^{eff,tot}_{9}=[2.793, 4.394 ]$, $C^{\prime~eff,tot}_{9}=[0,1.586]$, $C^{tot}_{10}=[-5.846, -4.260]$ and $C^{\prime~tot}_{10}=[-1.586,0]$ for the corresponding Wilson coefficients. Since the $ C^{eff(tot)}_9 $ depend on $ q^2 $, the above intervals for these coefficients denote the maximum and minimum values obtained varying  $ q^2 $ in the physical region, i.e., $ [0- 20]$ $ GeV^{2} $. In the case of  coefficients with  label $ ``tot" $ the above  intervals  are obtained considering the intervals for related parameters in Eq. (\ref{constraints}). Note that we will use  directly the expressions of the Wilson coefficients in the numerical analyses instead of the above-mentioned values/intervals. We shall remark that the above mentioned values/intervals for $ C^{eff}_{7} $, $ C^{eff}_{9} $ and $ C_{10} $ are  consistent with the ones obtained in \cite{Descotes-Genon:2013wba,Hurth:2014vma,Beaujean:2015gba,Du:2015tda,Descotes-Genon:2015uva,Hurth:2016fbr} for Wilson coefficients using the global fits to $b \rar s \ell^+ \ell^- $ data. We would also  like to compare the intervals for  four Wilson coefficients $C^{eff,tot}_{9} $, $C^{\prime~eff,tot}_{9}$,  $C^{tot}_{10} $   and $ C^{\prime~tot}_{10} $, which are relevant  to the LQ model  with the values  extracted in \cite{Meinel:2016grj}  from experimental data on observables of 
$\Lambda_{b}\rightarrow \Lambda \mu^{+} \mu^{-}$ in a $(9,10,9^{'},10^{'})  $ scenario assuming uncorrelated independent contributions to these coefficients. In Ref. \cite{Meinel:2016grj} the values  $C_{9}= 6.0^{+0.8}_{-0.8}$, $C^{\prime}_{9} =0.5^{+1.3}_{-1.8}$,  $C_{10}=-1.3^{+1.3}_{-1.1} $   and $ C^{\prime}_{10}=2.3^{+0.8}_{-1.3} $ are obtained. The comparison of the intervals obtained in the present study with those of  Ref. \cite{Meinel:2016grj} shows that our prediction on the range of $C^{\prime}_{9}$ exactly remains inside the interval obtained in Ref. \cite{Meinel:2016grj}. For other coefficients although the values obtained in these works are comparable in some regions, we overall see  considerable differences between the predictions of two studies.  The difference in $C_{9}$ can be attributed to the fact that in Ref. \cite{Meinel:2016grj} the authors use the data only in the interval $ q^2=[15-20] $ $GeV^2$ to extract its value.

As we previously said, we  use the values of form factors calculated via
 light cone QCD sum rules in full theory and available for all channels under consideration from Refs. \cite{Aliev:2010uy,Azizi:2011if,Azizi:2011mw}. These form factors are also available in lattice QCD in $ \Lambda $ channel \cite{Detmold:2016pkz}.
\begin{table}
\begin{tabular}{|c|c|}
\hline\hline
Some Input Parameters & Values \\ \hline\hline
$ m_{\Lambda_b} $    &   $ 5.6195               $   $GeV$ \\
$ m_{\Lambda} $      &   $ 1.11568              $   $GeV$ \\
$ \tau_{\Lambda_b} $ &   $ 1.451\times 10^{-12} $   $s$      \\
$ m_{\Sigma_b} $    &   $ 5.807               $   $GeV$ \\
$ m_{\Sigma} $      &   $ 1.192             $   $GeV$ \\
$ \tau_{\Sigma_b} $ &   $ 1.391\times 10^{-12} $   $s$      \\
$ m_{\Xi_b} $    &   $ 5.791               $   $GeV$ \\
$ m_{\Xi} $      &   $ 1.314             $   $GeV$ \\
$ \tau_{\Xi_b} $ &   $ 1.464\times 10^{-12} $   $s$      \\
$ m_W $              &   $ 80.385               $   $GeV$ \\
$ G_{F} $            &   $ 1.166\times 10^{-5}  $   $GeV^{-2}$ \\
$ \alpha_{em} $      &   $ 1/137                $               \\
$ | V_{tb}V_{ts}^*|$ &   $ 0.040                $                \\
 \hline \hline
\end{tabular}%
\caption{The values of some input parameters  used in our analysis \cite{pdg}.}
\label{tab:Param}
\end{table}
\begin{table}
\begin{tabular}{|c|c|}
\hline\hline
Quarks & masses in $\overline{MS}$ scheme \\ \hline\hline
$ m_c $              &   $ (1.275\pm0.025)       $   $GeV$ \\
$ m_b $              &   $ (4.18\pm0.03)           $   $GeV$ \\
$ m_t $              &   $ 160^{+4.8}_{-4.3}             $   $GeV$ \\
 \hline \hline
\end{tabular}%
\caption{The values of quark masses in $\overline{MS}$ scheme \cite{pdg}.}
\label{tab:Param}
\end{table}
The differential branching ratios of decay channels under consideration on $q^2$, in the SM and LQ models, at $\mu$ and $\tau$ lepton channels are plotted in Figures 1-6.
Note that, in these figures, the form factors are encountered with their uncertainties in both models.  The bands in LQ
model are due to both the constrained regions of some parameters presented in Eq. (\ref{constraints}) and errors of form factors. In these figures, we  show  the charmonia veto regions by the  vertical shaded bands.
  We do not present the results for $ e $ channel in the figures, because the predictions of $ \mu $ channel are very close to those of the $ e $  channel.
 In figure 1, we also show the experimental data provided by LHCb \cite{LHCb} and lattice predictions \cite{Detmold:2016pkz}.
 From these figures  it is clear that,
\begin{itemize}
\item the bands of  differential branching ratios in terms of $q^2$ obtained in SM  for all baryonic processes at both  lepton channels remain inside the bands of the  LQ model. The LQ model bands are wider and somewhere show considerable  discrepancies  from the SM predictions for all channels roughly at whole physical regions of  $q^2$.
\item The  SM  band for the differential branching fraction in   $\Lambda_{b}\rightarrow \Lambda \mu^{+} \mu^{-}$ channel  roughly coincides with  all the  lattice predictions borrowed from Ref. \cite{Detmold:2016pkz}. This band also defines  all the experimental data  provided by the LHCb Collaboration except  that in 
the interval   $18$ GeV$^2/$c$^4$ $\leq q^2\leq 20$ GeV$^2/$c$^4$, which can not be described by the SM. This datum coincides with the LQ band. As is also seen from this figure   the lattice QCD predictions on the differential branching fraction in   $\Lambda_{b}\rightarrow \Lambda \mu^{+} \mu^{-}$ channel  show considerable discrepancies with the experimental data in the interval  $15$ GeV$^2/$c$^4$ $\leq q^2\leq 20$ GeV$^2/$c$^4$.
\end{itemize}
\begin{figure}[h!]
\centering
\begin{tabular}{cc}
\epsfig{file=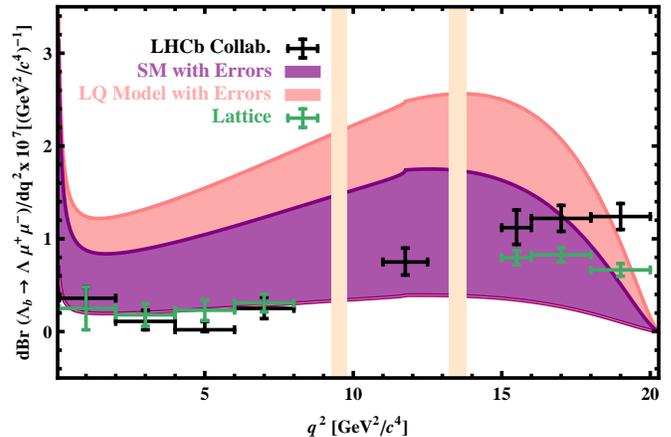,width=1.0\linewidth,clip=} &
\end{tabular}
\caption{The dependence of the differential branching ratio on  $q^2$  for the $\Lambda_{b}\rightarrow \Lambda \mu^{+} \mu^{-}$  transition in the SM and LQ models. 
The experimental data are taken from the LHCb Collaboration  Ref. \cite{LHCb}.  The lattice predictions are borrowed from Ref. \cite{Detmold:2016pkz}.The vertical shaded bands indicate the charmonia veto regions.}
\end{figure}
\begin{figure}[h!]
\centering
\begin{tabular}{cc}
\epsfig{file=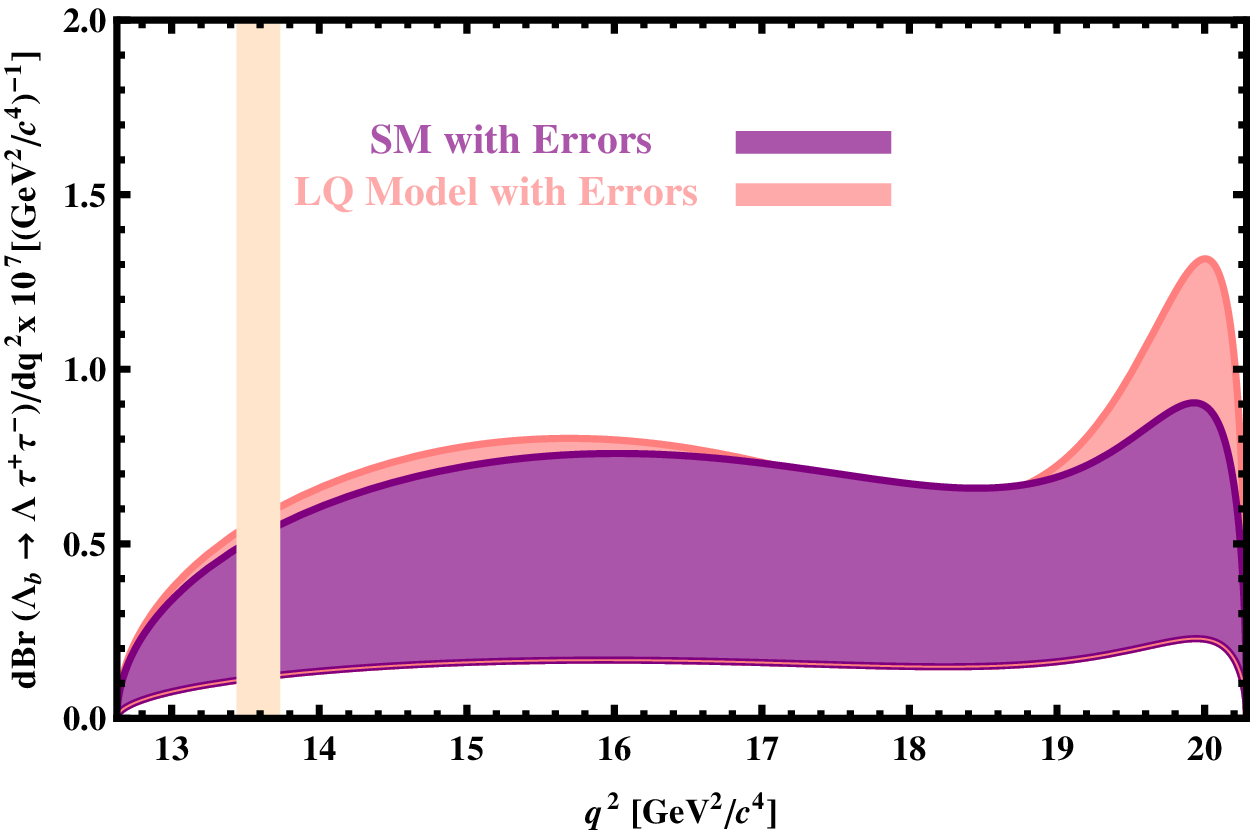,width=1.0\linewidth,clip=} &
\end{tabular}
\caption{The dependence of the differential branching ratio on  $q^2$  for the $\Lambda_{b}\rightarrow \Lambda \tau^{+} \tau^{-}$ transition in the SM and LQ models. The vertical shaded band indicates the charmonia veto region.}
\end{figure}

\begin{figure}[h!]
\centering
\begin{tabular}{cc}
\epsfig{file=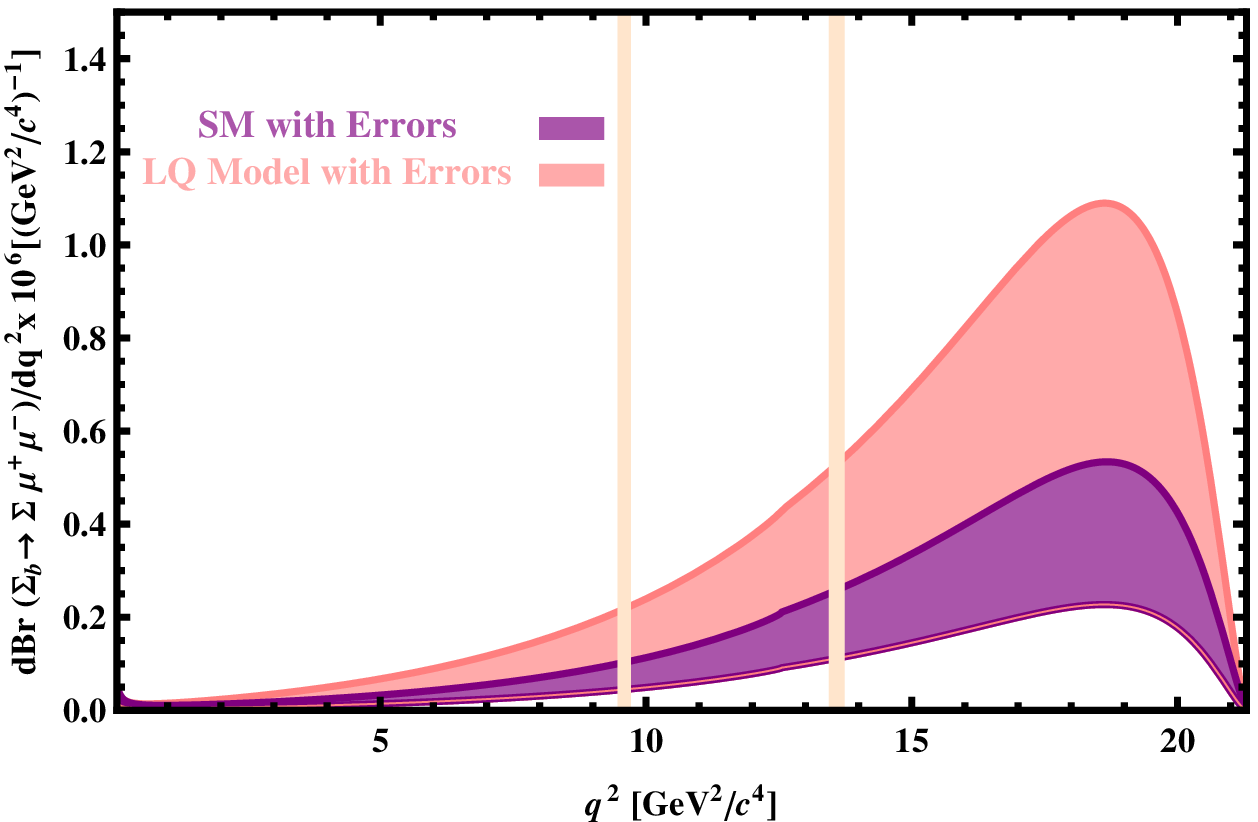,width=1.0\linewidth,clip=} &
\end{tabular}
\caption{The dependence of the differential branching ratio on  $q^2$  for the $\Sigma_{b}\rightarrow \Sigma \mu^{+} \mu^{-}$ transition in the SM and LQ models. The vertical shaded bands indicate the charmonia veto regions.}
\end{figure}
\begin{figure}[h!]
\centering
\begin{tabular}{cc}
\epsfig{file=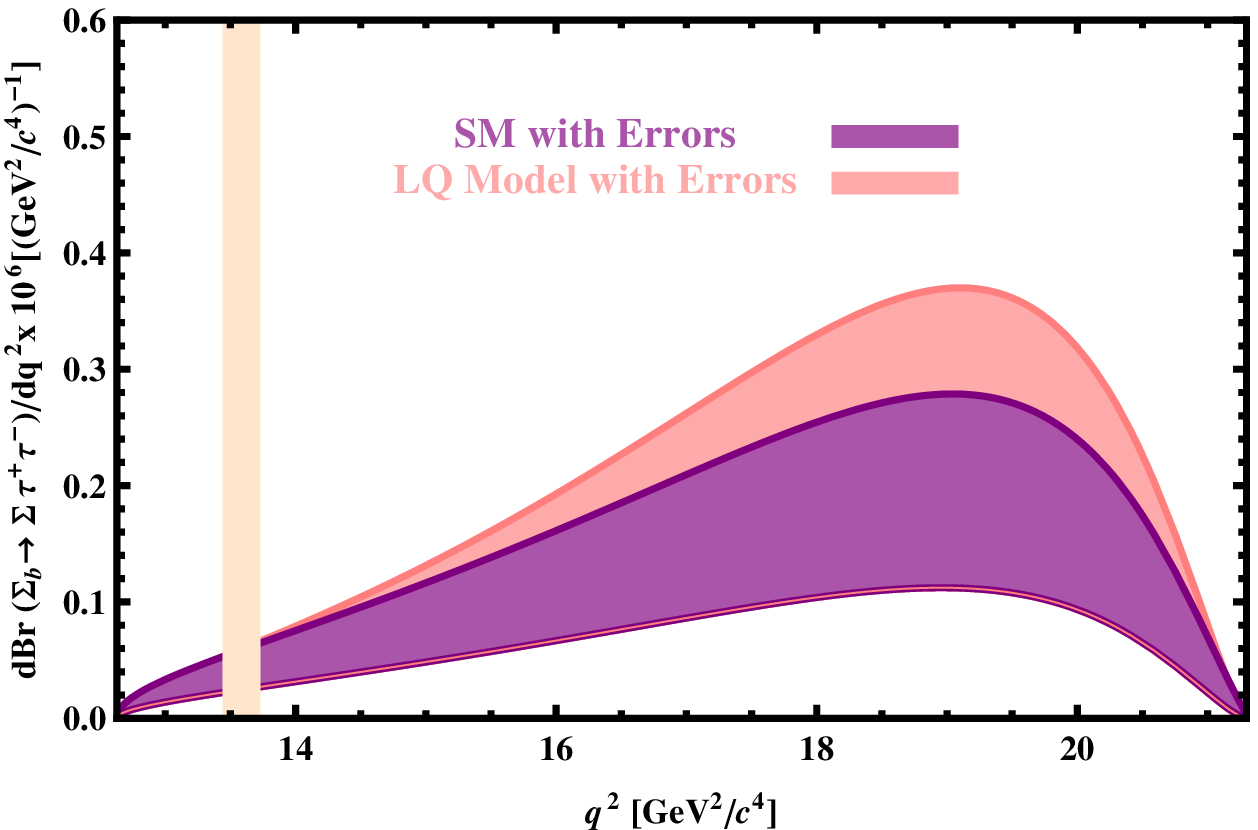,width=1.0\linewidth,clip=} &
\end{tabular}
\caption{The dependence of the differential branching ratio on  $q^2$  for the $\Sigma_{b}\rightarrow \Sigma \tau^{+} \tau^{-}$ transition in the SM and LQ models. The vertical shaded band indicates the charmonia veto region.}
\end{figure}

\begin{figure}[h!]
\centering
\begin{tabular}{cc}
\epsfig{file=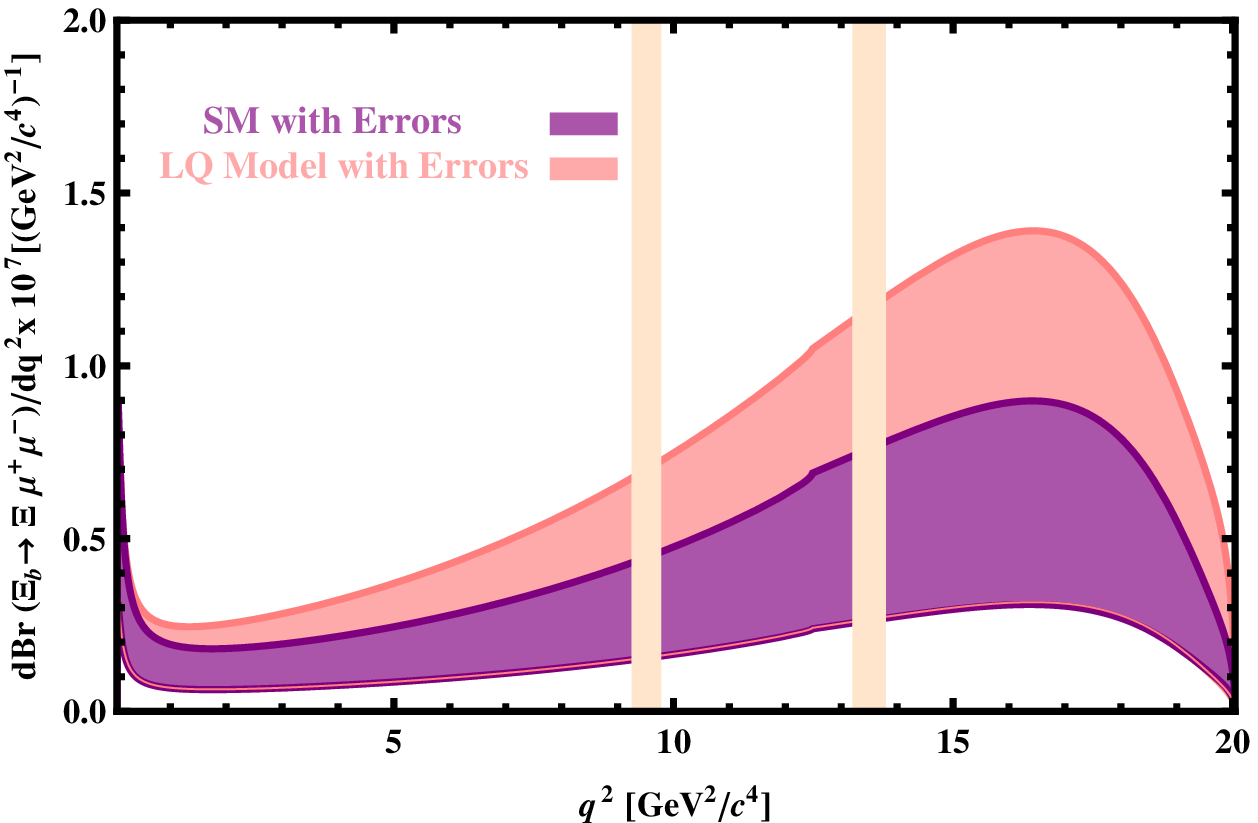,width=1.0\linewidth,clip=} &
\end{tabular}
\caption{The dependence of the differential branching ratio on  $q^2$  for the $\Xi_{b}\rightarrow \Xi \mu^{+} \mu^{-}$ transition in the SM and LQ models. The vertical shaded bands indicate the charmonia veto regions.}
\end{figure}
\begin{figure}[h!]
\centering
\begin{tabular}{cc}
\epsfig{file=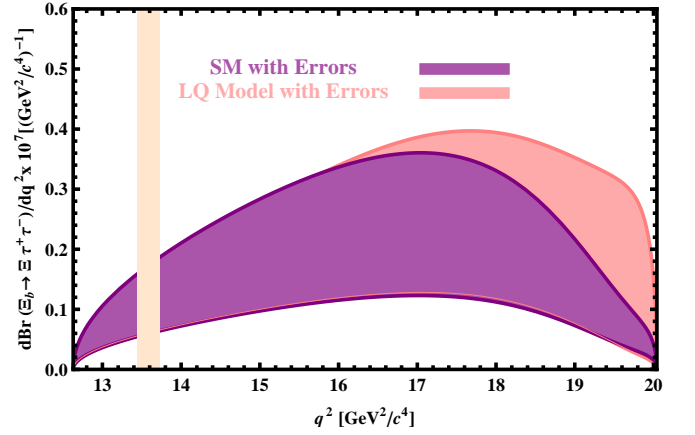,width=1.0\linewidth,clip=} &
\end{tabular}
\caption{The dependence of the differential branching ratio on  $q^2$  for the $\Xi_{b}\rightarrow \Xi \tau^{+} \tau^{-}$ transition in the SM and LQ models. The vertical shaded band indicates the charmonia veto region.}
\end{figure}

\subsection{The lepton forward-backward asymmetry}
In this subsection, we present the results of the lepton forward-backward asymmetry ($ {\cal A}_{FB}  $) which is one of useful observables to search for NP effects.
This quantity   is defined  as
\bea {\cal A}_{FB} (\hat s)=
\frac{\ds{\int_0^1\frac{d^{2}\Gamma}{d\hat{s}dz}}(z,\hat s)\,dz -
\ds{\int_{-1}^0\frac{d^{2}\Gamma}{d\hat{s}dz}}(z,\hat s)\,dz}
{\ds{\int_0^1\frac{d^{2}\Gamma}{d\hat{s}dz}}(z,\hat s)\,dz +
\ds{\int_{-1}^0\frac{d^{2}\Gamma}{d\hat{s}dz}}(z,\hat s)\,dz}~. 
\eea
\begin{figure}[h!]
\centering
\begin{tabular}{cc}
\epsfig{file=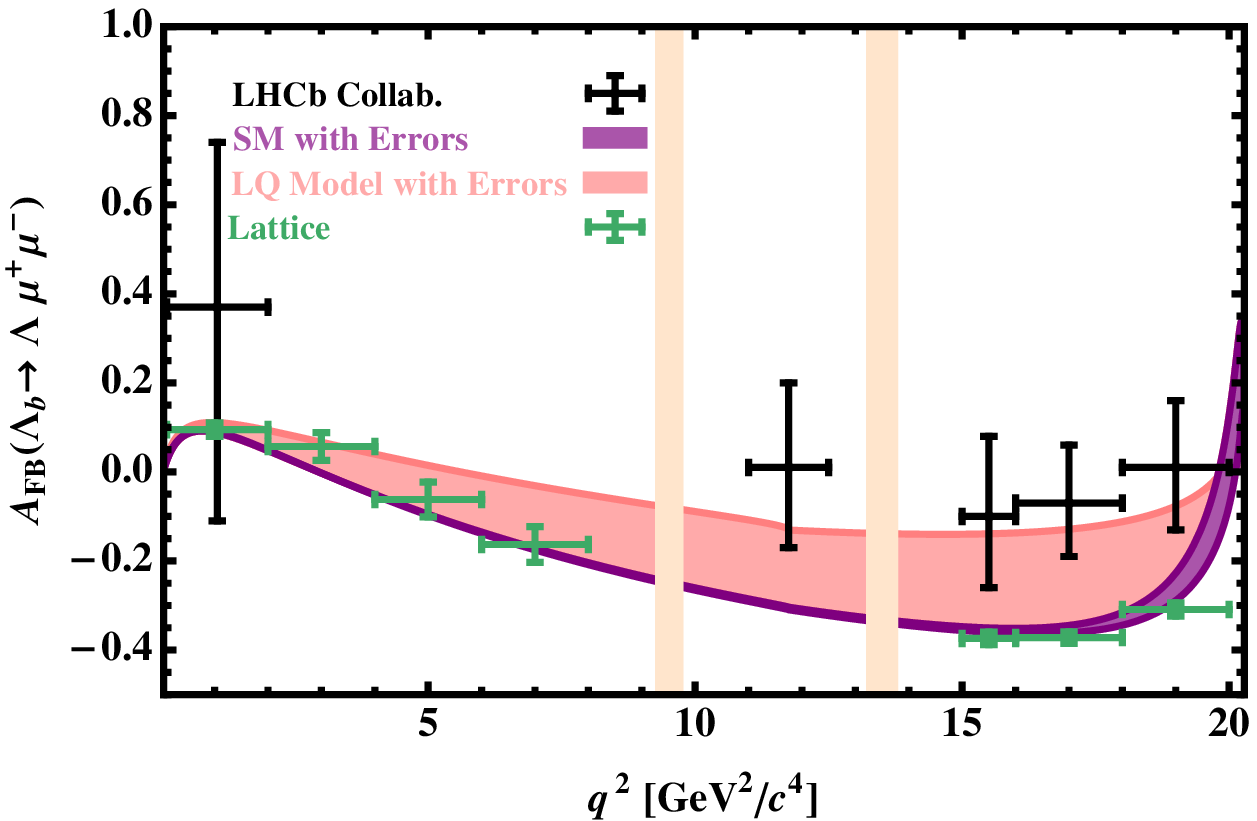,width=1.0\linewidth,clip=} &
\end{tabular}
\caption{The dependence of the $ {\cal A}_{FB}  $ on  $q^2$  for the $\Lambda_{b}\rightarrow \Lambda \mu^{+} \mu^{-}$ transition in the SM and 
LQ models. The experimental data are taken from the LHCb Collaboration Ref. \cite{LHCb}. The lattice predictions are borrowed from Ref. \cite{Detmold:2016pkz}. The vertical shaded bands indicate the charmonia veto regions.}
\end{figure}
\begin{figure}[h!]
\centering
\begin{tabular}{cc}
\epsfig{file=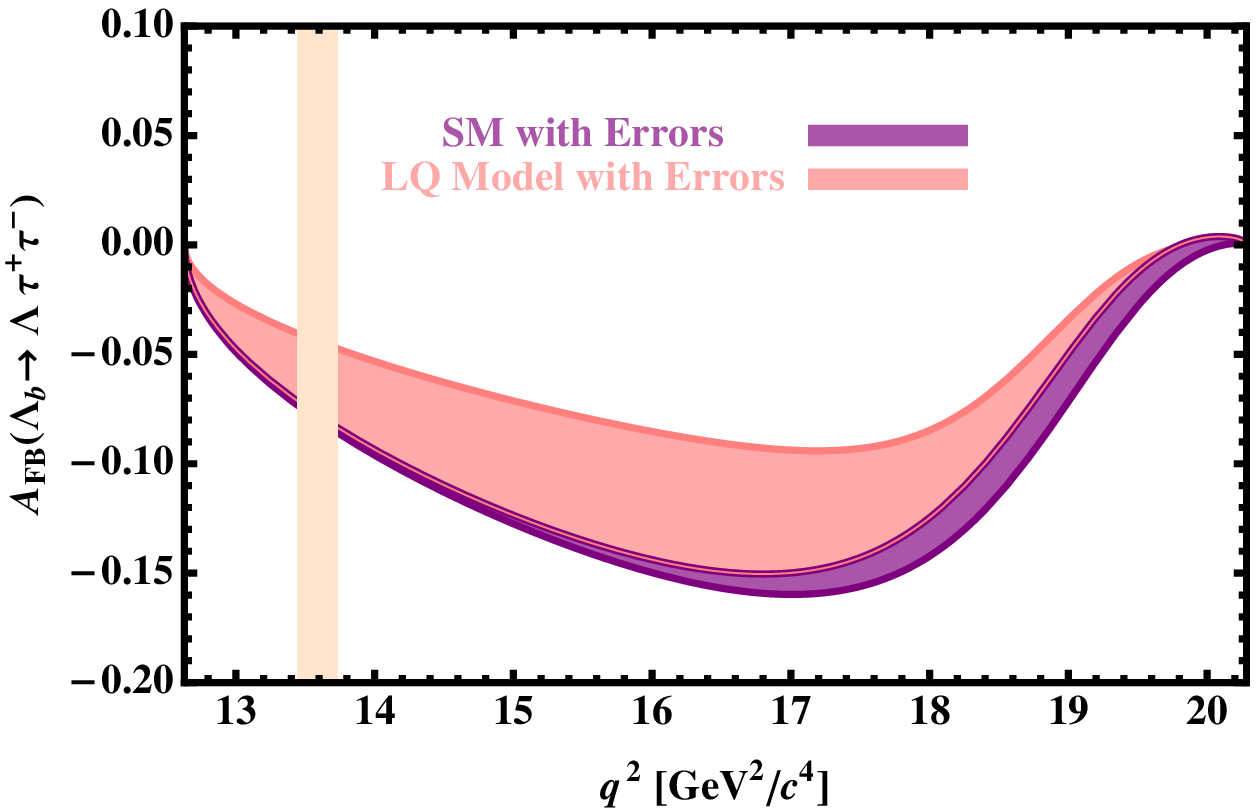,width=1.0\linewidth,clip=}&
\end{tabular}
\caption{The dependence of the $ {\cal A}_{FB}  $ on  $q^2$  for the $\Lambda_{b}\rightarrow \Lambda \tau^{+} \tau^{-}$ transition in the SM and 
LQ models. The vertical shaded band indicates the charmonia veto region.}
\end{figure}

\begin{figure}[h!]
\centering
\begin{tabular}{cc}
\epsfig{file=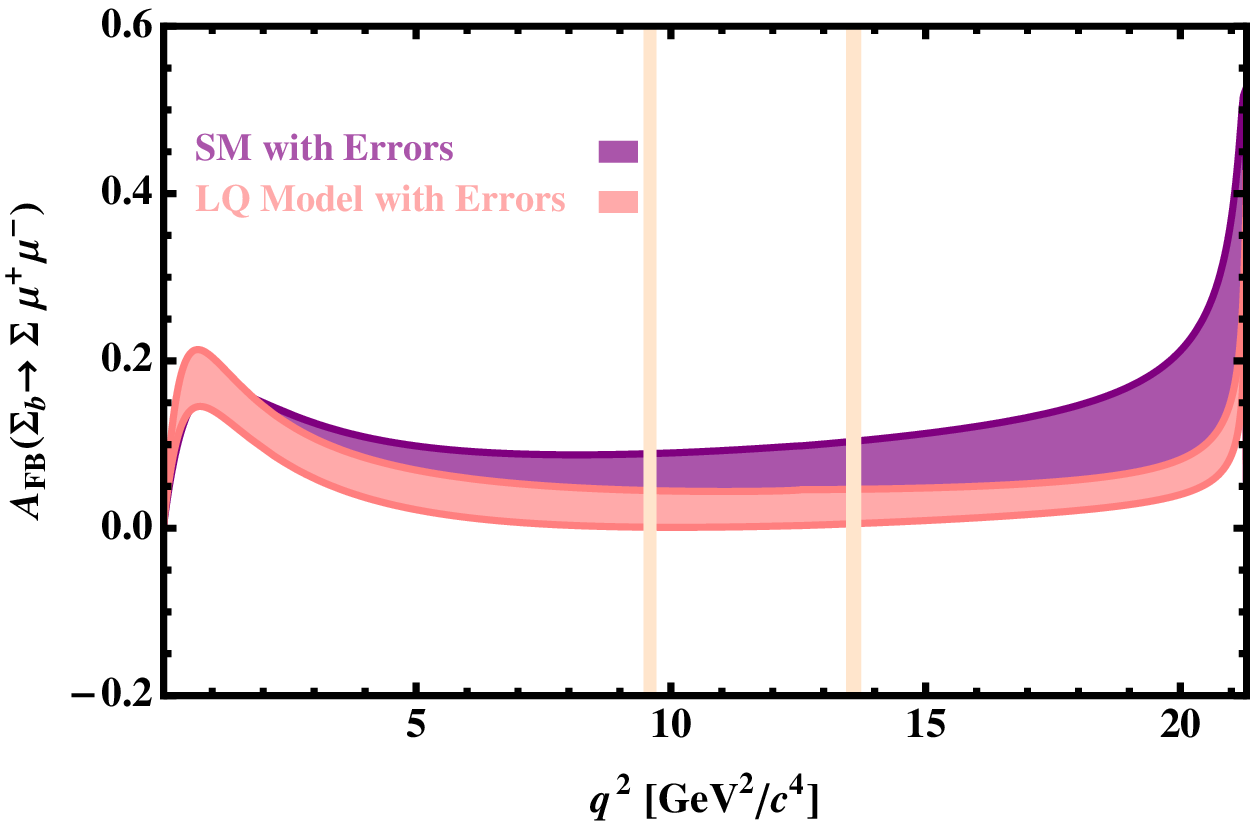,width=1.0\linewidth,clip=}&
\end{tabular}
\caption{The dependence of the $ {\cal A}_{FB}  $ on  $q^2$  for the $\Sigma_{b}\rightarrow \Sigma \mu^{+} \mu^{-}$ transition in the SM and LQ models. The vertical shaded bands indicate the charmonia veto regions.}
\end{figure}
\begin{figure}[h!]
\centering
\begin{tabular}{cc}
\epsfig{file=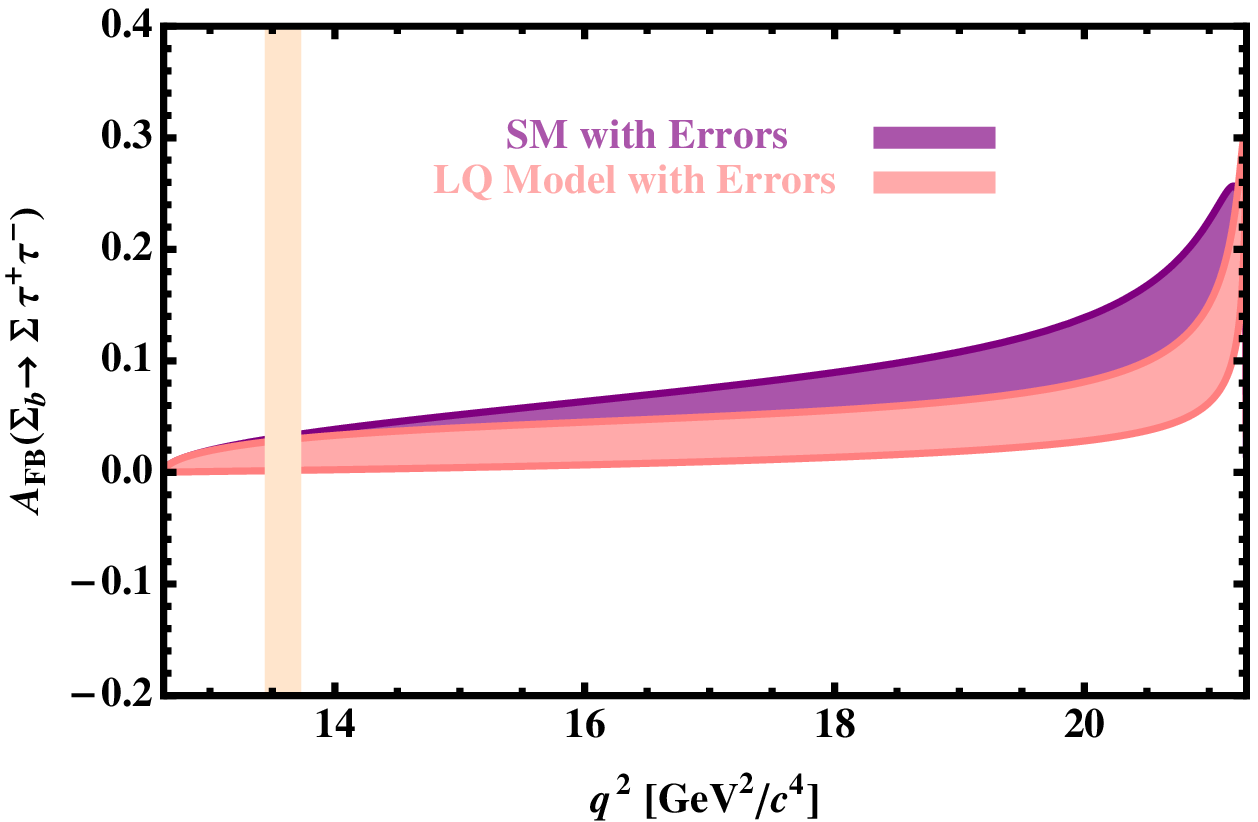,width=1.0\linewidth,clip=}& 
\end{tabular}
\caption{The dependence of the $ {\cal A}_{FB}  $ on  $q^2$  for the $\Sigma_{b}\rightarrow \Sigma \tau^{+} \tau^{-}$ transition in the SM and LQ models. The vertical shaded band indicates the charmonia veto region.}
\end{figure}

\begin{figure}[h!]
\centering
\begin{tabular}{cc}
\epsfig{file=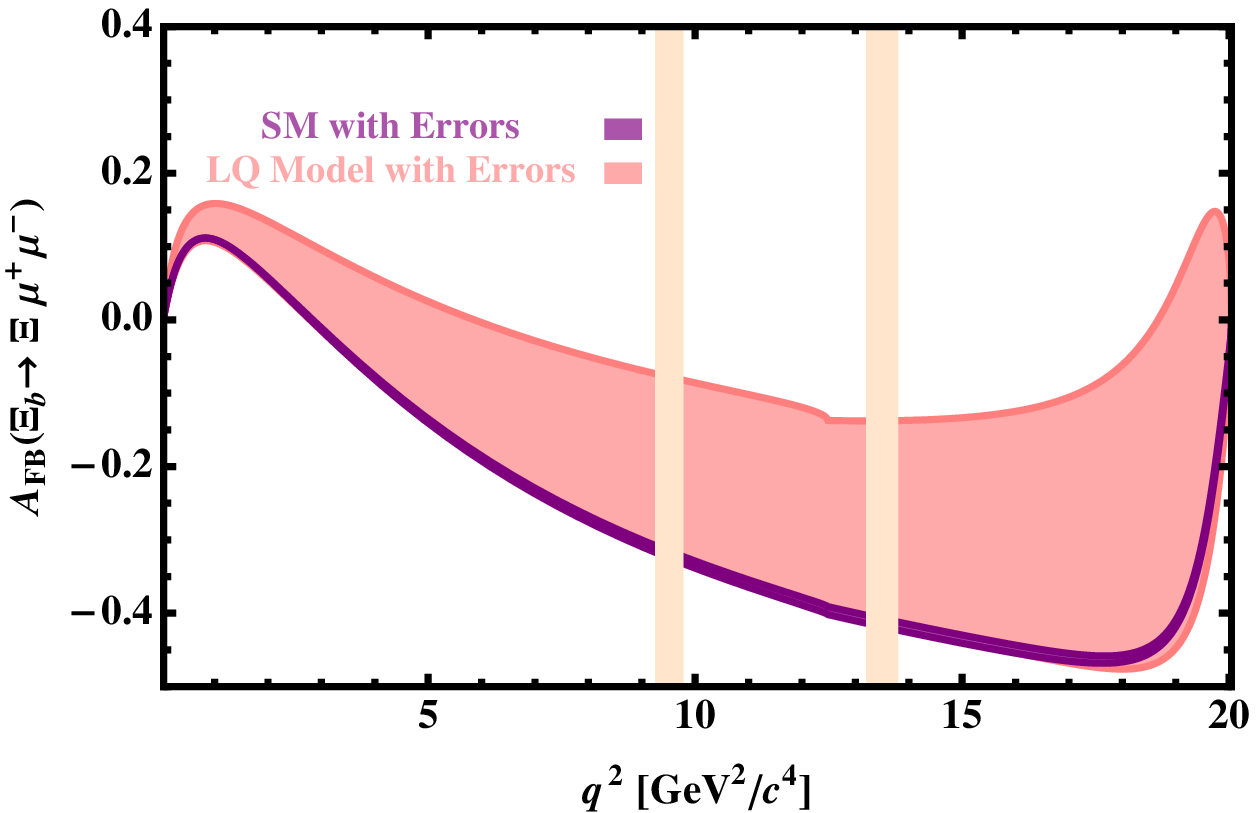,width=1.0\linewidth,clip=} &
\end{tabular}
\caption{The dependence of the $ {\cal A}_{FB}  $ on  $q^2$  for the $\Xi_{b}\rightarrow \Xi \mu^{+} \mu^{-}$ transition in the SM and LQ models. The vertical shaded bands indicate the charmonia veto regions.}
\end{figure}
\begin{figure}[h!]
\centering
\begin{tabular}{cc}
\epsfig{file=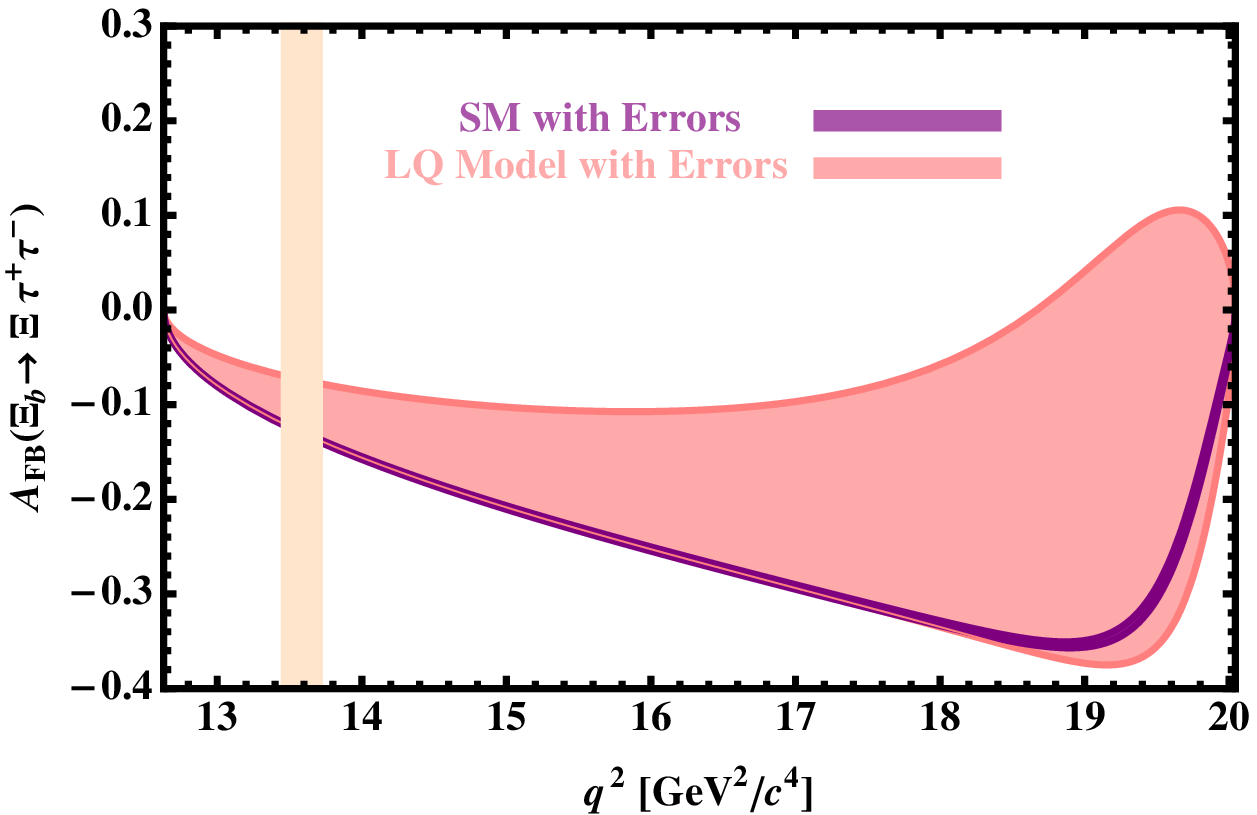,width=1.0\linewidth,clip=}&
\end{tabular}
\caption{The dependence of the $ {\cal A}_{FB}  $ on  $q^2$  for the $\Xi_{b}\rightarrow \Xi \tau^{+} \tau^{-}$ transition in the SM and LQ models. The vertical shaded band indicates the charmonia veto region.}
\end{figure}
In order to see how predictions of LQ scenario  deviate from those of the SM, we  plot the dependence of the lepton forward-backward asymmetry
  on  $q^2$ for the  channels under discussion  in Figures 7-12. In figure 7, we also present the measured values of the leptonic 
forward backward-asymmetries  by the  LHCb Collaboration  \cite{LHCb} as well as the lattice QCD predictions \cite{Detmold:2016pkz} in the $\Lambda_{b}\rightarrow \Lambda \mu^{+} \mu^{-}$ decay channel.
 From these figures, we read that
\begin{itemize}
\item in all decay channels the LQ model predictions demonstrate considerable discrepancies from the SM predictions.
\item The SM band on the lepton forward-backward asymmetry in  $\Lambda_{b}\rightarrow \Lambda \mu^{+} \mu^{-}$ channel coincides with the existing  lattice QCD predictions borrowed from Ref. \cite{Detmold:2016pkz}.
\item Ignoring from the small intersection of the SM narrow bands with errors of the experimental data at very low and high values of $q^2$, the LQ model, against the SM, can describe all data available in  $\Lambda_{b}\rightarrow \Lambda \mu^{+} \mu^{-}$ channel. The lattice QCD predictions in this channel also show sizable differences with the experimental data.
\end{itemize}

\section{Conclusion}
In the present work, we have performed a comprehensive analysis of the semileptonic $ \Lambda_b \rightarrow \Lambda \ell^+  \ell^-$, $\Sigma_b  \rightarrow \Sigma \ell^+  \ell^-$ and $\Xi_b \rightarrow \Xi \ell^+  \ell^-$ rare  processes in the SM as well as the scalar leptoquark model.
 Using the parametrization of the matrix elements  in terms of  form factors calculated via light cone QCD sum rules in the full theory,  we calculated the differential decay width
and numerically analyzed the differential branching fraction and the lepton forward-backward asymmetry in terms of $q^2$ in different heavy baryonic decay channels for both the $\mu$ and $\tau$ leptons 
in  both   scenarios. We compared the predictions of the LQ model on the considered physical observables with those of the SM and the existing  lattice QCD predictions as well as experimental data in $\Lambda_b  \rightarrow \Lambda \mu^+  \mu^-$
channel. We observed that the predictions of the LQ model  in all channels show considerable discrepancies with those of the SM on both the differential decay width and lepton forward-backward asymmetry. The SM results for both the observables considered in the present study are consistent with the existing predictions of lattice QCD. 
Except the interval  $18$ GeV$^2/$c$^4$ $\leq q^2\leq 20$ GeV$^2/$c$^4$, the SM band describes the existing  experimental data on the differential branching ratio in $\Lambda_b  \rightarrow \Lambda \mu^+  \mu^-$ transition. The datum 
in $18$ GeV$^2/$c$^4$ $\leq q^2\leq 20$ GeV$^2/$c$^4$ coincides with the LQ model prediction.


 In the case of lepton forward-backward asymmetry, the SM, overall, can not  describe the  experimental data existing in  $\Lambda_{b}\rightarrow \Lambda \mu^{+} \mu^{-}$  channel, while  the LQ model band coincides with the experimental data. 

More experimental data in $\Lambda_{b}\rightarrow \Lambda \tau^{+} \tau^{-}$  as well as  $\Sigma_b  \rightarrow \Sigma \ell^+  \ell^-$ and $\Xi_b \rightarrow \Xi \ell^+  \ell^-$ with both leptons are needed to compare with the
theoretical predictions. We  hope, with the RUN II data,  it will be possible to measure different physical quantities related to such FCNC transitions at LHCb in near future. Comparison of the future experimental
data with the theoretical predictions on different physical quantities in various  decay channels can help us better explain some anomalies between the SM predictions and the experimental data.
Any sizable discrepancy between  the theoretical predictions on physical observables with the experimental data can   be considered as an indication of new physics effects and   may  help us in the course of  searching for the new particles like leptoquarks.

\underline{Note Added}: \textit{ When preparing this work we noticed that  a part of our work, namely the $ \Lambda_b \rightarrow \Lambda \ell^+  \ell^-$ channel has been investigated
 in \cite{Sahoo:2016nvx,Wang:2016dne}  within the same framework. In  these studies the authors use the form factors, as the main inputs,  calculated in heavy
 quark effective theory while we use the form factors calculated via light cone QCD sum rules in full theory.}
\label{sec:Num}

\section*{ACKNOWLEDGEMENTS}

K.~A. thanks  Do\v{g}u\c{s} University for the financial support through  the grant BAP
2015-16-D1-B04.\\

\section*{CONFLICT OF INTEREST}
The authors declare that there is no conflict of interest regarding the publication of this paper.

\appendix*
\section{A}
\renewcommand{\theequation}{\Alph{section}.\arabic{equation}}
The Wilson coefficient $C_7^{eff}$ in leading logarithm approximation in the SM is written by \cite{Buras:1993xp,Misiak,Buras:1994dj,Buras:1998raa}
\begin{eqnarray}
\label{wilson-C7eff} C_7^{eff}(\mu_b) &=&
\eta^{\frac{16}{23}} C_7(\mu_W) \nnb \\
&+& \frac{8}{3} \left(\eta^{\frac{14}{23}} -\eta^{\frac{16}{23}} \right) C_8(\mu_W)\nnb \\
&+&C_2 (\mu_W) \sum_{i=1}^8 h_i \eta^{a_i}~, \nnb\\ 
\end{eqnarray}
where
 \begin{eqnarray} 
C_7(\mu_W) &=& -\frac{1}{2} D^{\prime}_{0}(x_t)~, \nnb \\
C_8(\mu_W) &=& -\frac{1}{2} E^{\prime}_{0}(x_t)~, \nnb \\
C_2(\mu_W) &=& 1~. 
\end{eqnarray} 
The functions $D^{\prime}_{0}(x_t)$ and $E^{\prime}_{0}(x_t)$ with $x_t=\frac{m_{t}^{2}}{m_{W}^{2}}$ are given as 
\begin{eqnarray} \label{Dprime0SM} 
D^{\prime}_{0}(x_t) &=& -\frac{(8 x_t^3+5 x_t^2-7 x_t)}{12 (1-x_t)^3} \nnb \\
&+& \frac{x_t^2(2-3 x_t)}{2(1-x_t)^4}\ln x_t~, \nnb \\
\end{eqnarray} 
\begin{eqnarray} \label{Eprime0SM} E^{\prime}_{0}(x_t) &=& - \frac{x_t(x_t^2-5 x_t-2)}{4(1-x_t)^3} \nnb \\ 
&+& \frac{3 x_t^2}{2 (1-x_t)^4}\ln x_t~. 
\end{eqnarray} 
The parameter $\eta$ in Eq.\eqref{wilson-C7eff} is defined as
\begin{eqnarray} \eta \es
\frac{\alpha_s(\mu_W)} {\alpha_s(\mu_b)}~,
\end{eqnarray}
with
\begin{eqnarray}
\alpha_s(x)=\frac{\alpha_s(m_Z)}{1-\beta_0\frac{\alpha_s(m_Z)}{2\pi}\ln(\frac{m_Z}{x})}~,
\end{eqnarray}
where $\alpha_s(m_Z)=0.118$ and $\beta_0=\frac{23}{3}$. The coefficients $h_i$ and $a_i$ in Eq.\eqref{wilson-C7eff} are also
written by \cite{Misiak,Buras:1994dj}
\be\frac{}{}
   \label{hi}
\begin{array}{rrrrrrrrrl}
h_i &=& (\!\! & 2.2996, & - 1.0880, & - \f{3}{7}, & - \f{1}{14}, & \nnb \\
&-& 0.6494, & -0.0380, & -0.0186, & -0.0057 & \!\!)  \vspace{0.1cm}, 
\end{array}
\ee 
\be\frac{}{}
   \label{ai}
\begin{array}{rrrrrrrrrl}
a_i &=& (\!\! & \f{14}{23}, & \f{16}{23}, & \f{6}{23}, & -\f{12}{23}, &  \\
&& 0.4086, & -0.4230, & -0.8994, & 0.1456 & \!\!).
\end{array}
\ee 
The Wilson coefficient $C_9^{eff}$ in SM is given by \cite{Misiak,Buras:1994dj}
\begin{eqnarray} \label{wilson-C9eff}
C_9^{eff}(\hat{s}^{\prime}) & = & C_9^{NDR}\eta(\hat s^{\prime}) \nnb \\
&+& h(z, \hat s^{\prime})\left( 3C_1 + C_2 + 3 C_3 + C_4 + 3C_5 + C_6 \right) \nonumber \\
& -& \f{1}{2} h(1, \hat s^{\prime}) \left( 4 C_3 + 4 C_4 + 3C_5 + C_6 \right) \nonumber \\
& -& \f{1}{2} h(0, \hat s^{\prime}) \left( C_3 + 3 C_4 \right) \nnb \\
&+& \f{2}{9} \left( 3 C_3 + C_4 + 3 C_5 +C_6 \right), 
\end{eqnarray}
where $\hat s^{\prime}=\frac{q^{2}}{m_{b}^{2}}$ with $4m_{l}^{2}\leq q^{2}\leq (m_{{\cal B}_{Q}}-m_{\cal B})^{2}$.
The $C_9^{NDR}$ in the naive dimensional regularization (NDR) scheme is written as
\begin{eqnarray} \label{C9NDR}C_9^{NDR} & = & P_0^{NDR} + \f{Y}{\sin^2\theta_W} \nnb \\
&-& 4 Z + P_E E, 
\end{eqnarray} 
where $P_0^{NDR}=2.60 \pm 0.25$, $\sin^2\theta_W=0.23$, $Y=0.98$ and $Z=0.679$
 \cite{Misiak,Buras:1994dj,Buras:1998raa}. The last term in Eq.\eqref{C9NDR} is ignored due to the negligible 
 value of $P_E$. In Eq.\eqref{wilson-C9eff}, the $\eta(\hat s^{\prime})$ is given as
\begin{eqnarray} 
\eta (\hat s^{\prime}) &=&  1 + \f{\alpha_{s}(\mu_b)}{\pi} \omega (\hat s^{\prime}), 
\end{eqnarray} 
with
\begin{eqnarray} \label{omega-shat}
\omega(\hat s^{\prime}) & = & - \f{2}{9} \pi^2 - \f{4}{3}\mbox{Li}_2(\hat s^{\prime}) - \f{2}{3}
\ln \hat s^{\prime} \ln(1-\hat s^{\prime})\nnb \\ 
&-& \f{5+4\hat s^{\prime}}{3(1+2\hat s^{\prime})}\ln(1-\hat s^{\prime}) \nonumber \\
& - &  \f{2 \hat s^{\prime} (1+\hat s^{\prime}) (1-2\hat s^{\prime})}{3(1-\hat s^{\prime})^2
(1+2\hat s^{\prime})} \ln \hat s^{\prime} \nnb \\ 
&+& \f{5+9\hat s^{\prime}-6\hat s^{\prime2}}
{6 (1-\hat s^{\prime}) (1+2\hat s^{\prime})}. 
\end{eqnarray}
The function $h(y,\hat s^{\prime})$ is written as
\begin{eqnarray} \label{h-phasespace} h(y,
\hat s^{\prime}) & = & -\f{8}{9}\ln\f{m_b}{\mu_b} - \f{8}{9}\ln y +
\f{8}{27} + \f{4}{9} x \nnb \\
& & - \f{2}{9} (2+x) |1-x|^{1/2} \nnb \\ 
&& \left\{
\begin{array}{ll}
\left( \ln\left| \f{\sqrt{1-x} + 1}{\sqrt{1-x} - 1}\right| - i\pi
\right), &
\mbox{for } x \equiv \f{4z^2}{\hat s^{\prime}} < 1 \nonumber \\
2 \arctan \f{1}{\sqrt{x-1}}, & \mbox{for } x \equiv \f {4z^2}{\hat
s^{\prime}} > 1,
\end{array}
\right. \\
\end{eqnarray}
where 
$y=1$ or $y=z=\frac{m_c}{m_b}$ 
and 
\begin{eqnarray} h(0, \hat s^{\prime})
& = & \f{8}{27} -\f{8}{9} \ln\f{m_b}{\mu_b} - \f{4}{9} \ln \hat s^{\prime}
+ \f{4}{9} i\pi.
\end{eqnarray}
The coefficients $C_j$ (j=1,...6) at $\mu_b=5~GeV$ scale are also written as \cite{Buras:1998raa}
\begin{eqnarray} \label{CJ} C_j=\sum_{i=1}^8 k_{ji}
\eta^{a_i} \qquad (j=1,...6) \vspace{0.2cm}, 
\end{eqnarray}
where the $k_{ji}$ are given as \be\frac{}{}
   \label{KJI}
\begin{array}{rrrrrrrrrl}
k_{1i} &=& (\!\! & 0, & 0, & \f{1}{2}, & - \f{1}{2}, & \nnb \\
&& 0, & 0, & 0, & 0 & \!\!),  \vspace{0.1cm} \\
k_{2i} &=& (\!\! & 0, & 0, & \f{1}{2}, &  \f{1}{2}, & \nnb \\
&& 0, & 0, & 0, & 0 & \!\!),  \vspace{0.1cm} \\
k_{3i} &=& (\!\! & 0, & 0, & - \f{1}{14}, &  \f{1}{6}, & \nnb \\
&& 0.0510, & - 0.1403, & - 0.0113, & 0.0054 & \!\!),  \vspace{0.1cm} \\
k_{4i} &=& (\!\! & 0, & 0, & - \f{1}{14}, &  - \f{1}{6}, & \nnb \\
&& 0.0984, & 0.1214, & 0.0156, & 0.0026 & \!\!),  \vspace{0.1cm} \\
k_{5i} &=& (\!\! & 0, & 0, & 0, &  0, & \nnb \\
&& - 0.0397, & 0.0117, & - 0.0025, & 0.0304 & \!\!) , \vspace{0.1cm} \\
k_{6i} &=& (\!\! & 0, & 0, & 0, &  0, & \nnb \\
&& 0.0335, & 0.0239, & - 0.0462, & -0.0112 & \!\!).  \vspace{0.1cm} \\
\end{array}
\ee

Considering the resonances from  $J/\psi$ family, we divide the  allowed physical region into the following three regions in the case of  the electron and  muon as final leptons: 
\begin{eqnarray*}
Region ~I   &;&4m_{l}^{2}\leq q^{2}\leq (m_{J/\psi(1s)}-0.02)^{2},\\
Region ~II &;&(m_{J/\psi(1s) }+0.02)^{2}\leq q^{2}\leq(m_{\psi(2s)}-0.02)^{2},\\
Region ~III &;&(m_{\psi(2s)}+0.02)^{2}\leq q^{2}\leq(m_{{\cal B}_{Q}}-m_{{\cal B}})^{2}.
\end{eqnarray*}%
In the case of  $\tau$, we have the following two regions:
\begin{eqnarray*}
Region ~I &;&4m_{\tau }^{2}\leq q^{2}\leq (m_{\psi(2s)}-0.02)^{2},\\
Region ~II &;&(m_{\psi(2s)}+0.02)^{2}\leq q^{2}\leq(m_{{\cal B}_{Q}}-m_{{\cal B}})^{2}.\\
\end{eqnarray*}%

The Wilson coefficient $C_{10}$ in the SM is given as:
\begin{eqnarray} \label{wilson-C10} 
C_{10}= - \frac{Y}{\sin^2 \theta_W}~. 
\end{eqnarray}
\section{B}
\renewcommand{\theequation}{\Alph{section}.\arabic{equation}}
\label{sec:App} 
The calligraphic coefficients used  in the transition amplitudes of the considered processes  are find as
\bea \label{coef-decay-rate} 
{\cal A}_{1} &=& f_1 C_{9}^{eff+} - g_1 C_{9}^{eff-} \nnb \\ 
& - & 2 m_b {1\over q^2} \Big[f_1^T C_{7}^{eff+} + 
g_1^T C_{7}^{eff-} \Big] ,~ \nnb \\
{\cal A}_{2} &=& {\cal A}_1 ( 1 \rar 2 ),~ \nnb \\ 
{\cal A}_{3} &=& {\cal A}_1 \ga 1 \rar 3 \dr ~,\nnb \\
{\cal B}_{1} &=& f_1 C_{9}^{eff+} + g_1 C_{9}^{eff-} \nnb \\ 
& - & 2 m_b {1\over q^2} \Big[ f_1^T C_{7}^{eff+} - 
g_1^T C_{7}^{eff-} \Big], ~ \nnb \\
 {\cal B}_{2} &=& {\cal B}_1 \ga 1 \rar 2 \dr, ~ \nnb \\
 {\cal B}_{3} &=& {\cal B}_1 \ga 1 \rar 3 \dr ~,\nnb \\
{\cal D}_{1} &=& f_1 C_{10}^{+} - g_1 C_{10}^{-},~ \nnb \\
 {\cal D}_{2} &=& {\cal D}_1 \ga 1 \rar 2 \dr~, \nnb \\
{\cal D}_{3} &=& {\cal D}_1 \ga 1 \rar 3 \dr~, \nnb \\
{\cal E}_{1} &=& f_1 C_{10}^{+} + g_1 C_{10}^{-}~, \nnb \\
{\cal E}_{2} &=& {\cal E}_1 \ga 1 \rar 2 \dr~, \nnb \\
{\cal E}_{3} &=& {\cal E}_1 \ga 1 \rar 3 \dr~, \nnb  \\
\eea
with
\bea
C_{9}^{eff+} & = & C_{9}^{eff}+C^{\prime~eff}_{9}~, \nnb \\
C_{9}^{eff-} & = & C_{9}^{eff}-C^{\prime~eff}_{9}~, \nnb \\
C_{7}^{eff+} & = & C_{7}^{eff}+C^{\prime~eff}_{7}~, \nnb \\ 
C_{7}^{eff-} & = & C_{7}^{eff}-C^{\prime~eff}_{7}~, \nnb \\
C_{10}^{+} & = & C_{10}+C^{\prime}_{10}~, \nnb \\ 
C_{10}^{-} & = & C_{10}-C^{\prime}_{10}~. \nnb
\eea
The functions
${\cal T}_{0}^{tot}(\hat s)$, ${\cal T}_{1}^{tot}(\hat s)$ and ${\cal T}_{2}^{tot}(\hat s)$  in the differential decay width are given as
\begin{widetext}
\bea {\cal T}_{0}^{tot}(\hat s) \es 32 m_\ell^2
m_{{\cal B}_Q}^4 \hat s (1+r-\hat s) \Big( \vel {\cal D}_{3} \ver^2 +
\vel {\cal E}_{3} \ver^2 \Big) \nnb \\
\ar 64 m_\ell^2 m_{{\cal B}_Q}^3 (1-r-\hat s) \, \mbox{\rm Re} \Big[{\cal D}_{1}^\ast
{\cal E}_{3} + {\cal D}_{3}
{\cal E}_1^\ast \Big] \nnb \\
\ar 64 m_{{\cal B}_Q}^2 \sqrt{r} (6 m_\ell^2 - m_{{\cal B}_Q}^2 \hat s)
{\rm Re} \Big[{\cal D}_{1}^\ast {\cal E}_{1}\Big] \nnb\\ 
\ar 64 m_\ell^2 m_{{\cal B}_Q}^3 \sqrt{r} \Bigg\{ 2 m_{{\cal B}_Q} \hat s
{\rm Re} \Big[{\cal D}_{3}^\ast {\cal E}_{3}\Big] + (1 - r + \hat s)
{\rm Re} \Big[{\cal D}_{1}^\ast {\cal D}_{3} + {\cal E}_{1}^\ast {\cal E}_{3}\Big]\Bigg\} \nnb \\
\ar 32 m_{{\cal B}_Q}^2 (2 m_\ell^2 + m_{{\cal B}_Q}^2 \hat s) \Bigg\{ (1
- r + \hat s) m_{{\cal B}_Q} \sqrt{r} \,
\mbox{\rm Re} \Big[{\cal A}_{1}^\ast {\cal A}_{2} + {\cal B}_{1}^\ast {\cal B}_{2}\Big] \nnb \\
\ek m_{{\cal B}_Q} (1 - r - \hat s) \, \mbox{\rm Re} \Big[{\cal A}_{1}^\ast {\cal B}_{2} +
{\cal A}_{2}^\ast {\cal B}_{1}\Big] - 2 \sqrt{r} \Big( \mbox{\rm Re} \Big[{\cal A}_{1}^\ast {\cal B}_{1}\Big] +
m_{{\cal B}_Q}^2 \hat s \,
\mbox{\rm Re} \Big[{\cal A}_{2}^\ast {\cal B}_{2}\Big] \Big) \Bigg\} \nnb \\
\ar 8 m_{{\cal B}_Q}^2 \Bigg\{ 4 m_\ell^2 (1 + r - \hat s) +
m_{{\cal B}_Q}^2 \Big[(1-r)^2 - \hat s^2 \Big]
\Bigg\} \Big( \vel {\cal A}_{1} \ver^2 +  \vel {\cal B}_{1} \ver^2 \Big) \nnb \\
\ar 8 m_{{\cal B}_Q}^4 \Bigg\{ 4 m_\ell^2 \Big[ \lambda + (1 + r -
\hat s) \hat s \Big] + m_{{\cal B}_Q}^2 \hat s \Big[(1-r)^2 - \hat s^2 \Big]
\Bigg\} \Big( \vel {\cal A}_{2} \ver^2 +  \vel {\cal B}_{2} \ver^2 \Big) \nnb \\
\ek 8 m_{{\cal B}_Q}^2 \Bigg\{ 4 m_\ell^2 (1 + r - \hat s) -
m_{{\cal B}_Q}^2 \Big[(1-r)^2 - \hat s^2 \Big]
\Bigg\} \Big( \vel {\cal D}_{1} \ver^2 +  \vel {\cal E}_{1} \ver^2 \Big) \nnb\\
\ar 8 m_{{\cal B}_Q}^5 \hat s v^2 \Bigg\{ - 8 m_{{\cal B}_Q} \hat s \sqrt{r}\,
\mbox{\rm Re} \Big[{\cal D}_{2}^\ast {\cal E}_{2}\Big] +
4 (1 - r + \hat s) \sqrt{r} \, \mbox{\rm Re}\Big[{\cal D}_{1}^\ast {\cal D}_{2}+{\cal E}_{1}^\ast {\cal E}_{2}\Big]\nnb \\
\ek 4 (1 - r - \hat s) \, \mbox{\rm Re}\Big[{\cal D}_{1}^\ast {\cal E}_{2}+{\cal D}_{2}^\ast {\cal E}_{1}\Big] +
m_{{\cal B}_Q} \Big[(1-r)^2 -\hat s^2\Big] \Big( \vel {\cal D}_{2} \ver^2 + \vel
{\cal E}_{2} \ver^2 \Big) \Bigg\}, \nnb \\
\eea
\bea {\cal T}_{1}^{tot}(\hat s) &=& -16 m_{{\cal B}_Q}^4\s1 v \sqrt{\lambda} 
\Bigg\{ 2 Re\Big({\cal A}_{1}^* {\cal D}_{1}\Big)-2Re\Big({\cal B}_{1}^* {\cal E}_{1}\Big)\nn\\
&+&2 m_{{\cal B}_Q} Re\Big({\cal B}_{1}^* {\cal D}_{2}-{\cal B}_{2}^* {\cal D}_{1}+{\cal A}_{2}^* {\cal E}_{1}
-{\cal A}_{1}^*{\cal E}_{2}\Big)\Bigg\}\nn\\
&+&32 m_{{\cal B}_Q}^5 \s1~ v \sqrt{\lambda} \Bigg\{
m_{{\cal B}_Q} (1-r)Re\Big({\cal A}_{2}^* {\cal D}_{2} -{\cal B}_{2}^* {\cal E}_{2}\Big)\nn\\
&+& \sqrt{r} Re\Big({\cal A}_{2}^* {\cal D}_{1}+{\cal A}_{1}^* {\cal D}_{2}-{\cal B}_{2}^*{\cal E}_{1}
-{\cal B}_{1}^* {\cal E}_{2}\Big)\Bigg\}, \nn\\
\eea\
and
\bea {\cal T}_{2}^{tot}(\hat s) \es - 8 m_{{\cal B}_Q}^4 v^2 \lambda \Big(\vel {\cal A}_{1} \ver^2 + \vel {\cal B}_{1} \ver^2 
+ \vel {\cal D}_{1} \ver^2 + \vel {\cal E}_{1} \ver^2 \Big) \nnb \\
\ar 8 m_{{\cal B}_Q}^6 \hat s v^2 \lambda \Big( \vel {\cal A}_{2} \ver^2 + \vel
{\cal B}_{2} \ver^2 + \vel {\cal D}_{2} \ver^2 + \vel {\cal E}_{2} \ver^2 \Big) ~.\nn\\ 
\eea\
\end{widetext}

\end{document}